\documentclass{aastex63}

\usepackage{color}

\received{}
\revised{}
\accepted{}
\submitjournal{ApJ}

\shorttitle{ALMA observations of the GMC in M33}
\shortauthors{Tokuda et al.}

\begin{document}

\title{ALMA Observations of Giant Molecular Clouds in M33 I: Resolving Star Formation Activities in the Giant Molecular Filaments Possibly Formed by a Spiral Shock}

\correspondingauthor{Kazuki Tokuda}
\email{tokuda@p.s.osakafu-u.ac.jp}

\author[0000-0002-2062-1600]{Kazuki Tokuda}
\affiliation{Department of Physical Science, Graduate School of Science, Osaka Prefecture University, 1-1 Gakuen-cho, Naka-ku, Sakai, Osaka 599-8531, Japan}
\affiliation{National Astronomical Observatory of Japan, National Institutes of Natural Science, 2-21-1 Osawa, Mitaka, Tokyo 181-8588, Japan}

\author{Kazuyuki Muraoka}
\affiliation{Department of Physical Science, Graduate School of Science, Osaka Prefecture University, 1-1 Gakuen-cho, Naka-ku, Sakai, Osaka 599-8531, Japan}

\author{Hiroshi Kondo}
\affiliation{Department of Physical Science, Graduate School of Science, Osaka Prefecture University, 1-1 Gakuen-cho, Naka-ku, Sakai, Osaka 599-8531, Japan}

\author{Atsushi Nishimura}
\affiliation{Department of Physical Science, Graduate School of Science, Osaka Prefecture University, 1-1 Gakuen-cho, Naka-ku, Sakai, Osaka 599-8531, Japan}

\author{Tomoka Tosaki}
\affiliation{Joetsu University of Education, Yamayashiki-machi, Joetsu, Niigata 943-8512, Japan}

\author{Sarolta Zahorecz}
\affiliation{Department of Physical Science, Graduate School of Science, Osaka Prefecture University, 1-1 Gakuen-cho, Naka-ku, Sakai, Osaka 599-8531, Japan}
\affiliation{National Astronomical Observatory of Japan, National Institutes of Natural Science, 2-21-1 Osawa, Mitaka, Tokyo 181-8588, Japan}

\author{Sachiko Onodera}
\affiliation{Meisei University, 2-1-1 Hodokubo, Hino, Tokyo 191-0042, Japan}

\author{Rie E. Miura}
\affiliation{National Astronomical Observatory of Japan, National Institutes of Natural Science, 2-21-1 Osawa, Mitaka, Tokyo 181-8588, Japan}

\author{Kazufumi Torii}
\affiliation{Nobeyama Radio Observatory, National Astronomical Observatory of Japan (NAOJ), National Institutes of Natural Sciences (NINS), 462-2 Nobeyama, Minamimaki, Minamisaku-gun, Nagano 384-1305, Japan}

\author{Nario Kuno}
\affiliation{Department of Physics, Graduate School of Pure and Applied Sciences, University of Tsukuba, 1-1-1 Tennodai, Tsukuba, Ibaraki 305-8577, Japan}
\affiliation{Tomonaga Center for the History of the Universe, University of Tsukuba, Tsukuba, Ibaraki 305-8571, Japan}

\author{Shinji Fujita}
\affiliation{Department of Physical Science, Graduate School of Science, Osaka Prefecture University, 1-1 Gakuen-cho, Naka-ku, Sakai, Osaka 599-8531, Japan}

\author{Hidetoshi Sano}
\affiliation{Department of Physics, Nagoya University, Chikusa-ku, Nagoya 464-8602, Japan}
\affiliation{Institute for Advanced Research, Nagoya University, Furo-cho, Chikusa-ku, Nagoya 464-8601, Japan}

\author{Toshikazu Onishi}
\affiliation{Department of Physical Science, Graduate School of Science, Osaka Prefecture University, 1-1 Gakuen-cho, Naka-ku, Sakai, Osaka 599-8531, Japan}

\author{Kazuya Saigo}
\affiliation{National Astronomical Observatory of Japan, National Institutes of Natural Science, 2-21-1 Osawa, Mitaka, Tokyo 181-8588, Japan}

\author{Yasuo Fukui}
\affiliation{Department of Physics, Nagoya University, Chikusa-ku, Nagoya 464-8602, Japan}
\affiliation{Institute for Advanced Research, Nagoya University, Furo-cho, Chikusa-ku, Nagoya 464-8601, Japan}

\author{Akiko Kawamura}
\affiliation{National Astronomical Observatory of Japan, National Institutes of Natural Science, 2-21-1 Osawa, Mitaka, Tokyo 181-8588, Japan}

\author{Kengo Tachihara}
\affiliation{Department of Physics, Nagoya University, Chikusa-ku, Nagoya 464-8602, Japan}

\begin{abstract}

We report molecular line and continuum observations toward one of the most massive giant molecular clouds (GMCs), GMC-16, in M33 using ALMA with an angular resolution of 0\farcs44\,$\times$\,0\farcs27 ($\sim$2\,pc\,$\times$\,1\,pc). We have found that the GMC is composed of several filamentary structures in $^{12}$CO and $^{13}$CO\,($J$\,=\,2--1). The typical length, width, and total mass are $\sim$50--70\,pc, $\sim$5--6\,pc, and $\sim$10$^{5}$\,$M_{\odot}$, respectively, which are consistent with those of giant molecular filaments (GMFs) as seen in the Galactic GMCs. The elongations of the GMFs are roughly perpendicular to the direction of the galaxy's rotation, and several H$\;${\sc ii} regions are located at the downstream side relative to the filaments with an offset of $\sim$10--20\,pc. These observational results indicate that the GMFs are considered to be produced by a galactic spiral shock. The 1.3\,mm continuum and C$^{18}$O\,($J$\,=\,2--1) observations detected a dense clump with the size of $\sim$2\,pc at the intersection of several filamentary clouds, which is referred to as the $``$hub filament,$"$ possibly formed by a cloud--cloud collision. A strong candidate for protostellar outflow in M33 has also been identified at the center of the clump. We have successfully resolved the parsec-scale local star formation activity in which the galactic scale kinematics may induce the formation of the parental filamentary clouds.

\end{abstract}

\keywords{stars: formation  --- stars: protostars --- ISM: clouds--- ISM:  kinematics and dynamics --- ISM: individual (M33-GMC-16) --- galaxies: Local Group}

\section{Introduction} \label{sec:intro}
Giant molecular clouds (GMCs) are considered to be major sites of high-mass star-formation \citep[e.g.,][]{Heyer15}, and the formed stars eventually regulate the galaxy evolution through their feedback. Understanding of the physical properties of GMCs and their evolution is a challenging study observationally. The sizes of GMCs are as large as $>$10--100\,pc, and a large number of samples are needed to track the continuous evolution. Several comprehensive molecular gas surveys have been carried out toward the Local Group of galaxies, such as the Large Magellanic Cloud (LMC; e.g.,  \citealt{Fukui99,Kawamura09,Wong11}) and M33 \citep[e.g.,][]{,Miura12,Corbelli17}. They classified the GMCs into a few groups with different evolutionary stages based on the association of H$\;${\sc ii} regions and young massive star clusters, and then estimated the GMC lifetime as a few $\times$ 10\,Myr.

Star formation and GMC evolution in the spiral arm are classically considered to be controlled by galactic shock \citep{Fujimoto68,Roberts69,Shu73} caused by $``$quasi-stationary density waves$"$ \citep[e.g.,][]{Lin64}. In contrast to this, an alternative model called $``$dynamic$"$ spiral theory, which involves nonsteady stellar arms, has been proposed (\cite{Dobbs14} and references therein). The two models predict qualitatively different gas distributions on the spiral arm. The former produces an apparent offset between parental gas and H$\;${\sc ii} regions, but the latter does not show a clear spatial offset between them \citep{Wada11}.
Although recent observations toward grand design spiral galaxies sometimes prefer to show a sequential distribution of gas and high-mass stars across the spiral arm predicted by galactic shock \citep[e.g.,][]{Egusa04,Egusa11,Hirota11,Colombo14}, the actual mechanism to trigger the star formation in the spiral arm is not well constrained by available observations. On spiral arms in galaxies, gas is exposed to many processes in addition to spiral shocks, such as cloud--cloud collision, hydrodynamic instabilities, stellar feedback, and self-gravity (see the review by \citealt{Dobbs14}). In order to understand the mechanism driving GMC and high-mass star formation in spiral arms, it is important to reveal a high-dynamic-range picture from the galactic scale down to filaments/clumps (see the next paragraph) directly leading to star formation.

Molecular gas surveys using CO and its isotopes toward star-forming regions in the solar neighborhood, such as Taurus, revealed that filamentary structures are considered to be fundamental ingredients of molecular clouds \citep[e.g.,][]{Mizuno95,Onishi96,Goldsmith08,Hacar13} and eventually collapse into individual dense cores and protostars. Recent high-resolution dust continuum observations with the Herschel telescope confirmed the quasi-universality of the filamentary structure in molecular clouds extending to the Galactic plane (\citealt{Andre14} and references therein), and the typical widths of the filaments were measured as $\sim$0.1\,pc \citep{Doris11,Doris19,Andre16}. Although the typical length of the abovementioned filaments in the solar-neighborhood star-forming regions are $\sim$10\,pc or less, the observations toward the Galactic plane identify much longer filamentary complexes with the length scale of $\gtrsim$50-100\,pc, called $``$giant molecular filaments (GMFs).$"$ One of the most prominent examples is the $``$Nessie$"$ Nebula \citep{Jackson10}, which is an infrared dark cloud on the Galactic plane. 
The long filamentary structure possibly formed by the passage of a spiral shock. One of the most active high-mass star-forming molecular complexes in the Galaxy, W51, also has a long filamentary stream with a length of $\sim$100\,pc \citep[e.g.,][]{Burton70,Moon98,Okumura01,Fujita19}. The velocity of this stream is different from that of the main complex of W51, and this excess of the velocity may be due to the streaming motions induced by a spiral density-wave \citep{Burton70,Koo99}. However, the actual origin of these GMFs/streamer may be hard to understand due to the serious contaminations at the line of sight in the Galactic plane and its edge-on view. 

ALMA observations with an angular resolution of 0\farcs25 ($\sim$0.06\,pc) toward GMCs in the LMC have started to resolve $\sim$0.1\,pc width molecular filaments \citep{Fukui19,Tokuda19}, possibly formed by a galactic scale gas flow. Moderate resolution ($\sim$3\arcsec = 0.7\,pc) studies in the LMC also provided us hints for understanding the evolution of internal structures of GMCs. For example, \cite{Sawada18} suggest that the quiescent GMC shows a diffuse emission, whereas the active star-forming ones have highly structured distributions, i.e., filaments and clumps. \cite{Wong19} measured the linewidth-size relation in several GMCs down to $\lesssim$1\,pc and show the velocity dispersion at a fixed size slightly increases as star formation progresses. 

The early millimeter/submillimeter observations described in this section suggest that we need at least $\sim$1\,pc resolution to understand the internal gas properties of molecular clouds and their kinematics. In addition to this, a birds-eye view is needed to understand the relation between spiral arms and local star formation activities. One of the closest galaxies, the LMC and Small Magellanic Cloud, are not the best targets in this purpose because the irregular galaxies do not have clear spiral arms. The flocculent spiral galaxy M33 is the most unique candidate so far to investigate the effect of the spiral arms on molecular cloud and high-mass star formation at a parsec-scale resolution using ALMA thanks to its proximity ($\sim$840 kpc; \citealt{Freedman01}) and favorable inclination ($i$ = 51\arcdeg; \citealt{Corbelli00}). 
 
We have performed ALMA observations with a spatial resolution of $\sim$2\,pc $\times$ 1\,pc toward three massive ($\sim$10$^6$\,$M_{\odot}$) GMCs (NGC~604-GMC, GMC-8, and GMC-16) in different evolutionary stages identified by the early surveys \citep{Roso07,Onodera10,Miura12} to investigate the molecular gas structures and star-formation activities. In this paper, we present the results of GMC-16 associated with several H$\;${\sc ii} regions and 24\,$\micron$ sources \citep{Verley07}. We note that CO\,(3--2) observations of M33 by \cite{Miura12} cataloged the present target as two clouds, GMC-2 and GMC-16, which correspond to the cloud numbers of 425 and 435 in \cite{Corbelli17}, respectively. Because the two clouds spatially connect each other, we treat these objects as a single molecular cloud system for convenience throughout the manuscript and use the name of GMC-16 as a representative, which is more luminous in CO\,(3--2) in CO(3-2) than the other. We describe the detailed results of the other targets in a separate paper for NGC~604-GMC (K. Muraoka et al. 2020, in preparation) and a forthcoming paper for GMC-8.

\section{Observations} \label{sec:obs}
We performed ALMA Cycle\,5 Band\,6 (1.3\,mm) observations in molecular lines and continuum toward three GMCs in M33 (P.I.: K., Muraoka, \#2017.1.00461.S). We used the ALMA main array (the 12\,m array) with the configuration of C43-5 as well as the 7\,m array of the Atacama Compact Array (ACA; a.k.a. Morita Array). The observations were carried out during 2017 October and 2018 October. There were three spectral windows targeting $^{12}$CO\,($J$\,=\,2--1), $^{13}$CO\,($J$\,=\,2--1), and C$^{18}$O\,($J$\,=\,2--1). The bandwidths of the correlator setting were 117.19\,MHz with 1920 channels for $^{12}$CO and 960 channels for $^{13}$CO/C$^{18}$O. We used two spectral windows for the continuum observations with an aggregate bandwidth of 3750\,MHz with a channel width of 0.98\,MHz. 

While we did not change the system calibration provided by the observatory, the data was reprocessed with the Common Astronomy Software Application (CASA) package \citep{McMullin07} version 5.4.0 in the imaging process. We used the \texttt{tclean} task with the \texttt{multi-scale} deconvolver. We applied the Briggs weighting with a robust parameter of 0.5 and the natural weighting to the 12\,m and 7\,m array data, respectively. We used the \texttt{auto-multithresh} procedure \citep{Kepley20} in \texttt{tclean} to select the emission mask in the dirty and residual images. We continued the deconvolution process until the intensity of the residual image reached the $\sim$1$\sigma$ noise level.
We combined the individually imaged 12 and 7\,m array data sets with the \texttt{feathering} task. We also performed the multiple array data combination with an alternative method, in which the visibility data of the 12 and 7\,m array were merged together before the \texttt{tclean} task, and then imaged them using the same auto-masking technique as described above. The two different methods reproduce fairly similar results (only $\lesssim$10\% flux difference in rms), and we used the first option in the analysis through the manuscript. 

The final processed image qualities are summarized as follows. The beam sizes of the molecular lines ($^{12}$CO, $^{13}$CO, and C$^{18}$O) and continuum data are 0\farcs44\,$\times$\,0\farcs27, corresponding to $\sim$2\,pc\,$\times$\,1\,pc, and 0\farcs40\,$\times$\,0\farcs25, respectively. The rms noise levels of the molecular lines at a velocity resolution of $\sim$0.2\,km\,s$^{-1}$ are $\sim$4.3\,mJy\,beam$^{-1}$ ($\sim$0.9\,K). The sensitivity of the continuum observations is $\sim$0.02\,mJy\,beam$^{-1}$.

We estimated the missing flux of the interferometric observations toward GMC-16 using the available $^{12}$CO\,($J$\,=\,2--1) data obtained with the single-dish IRAM\,30\,m telescope \citep{Druard14}. 
We spatially smoothed the 12\,m+7\,m data to an angular resolution of 12\arcsec, which is the same as the IRAM data and then compared the total flux between the two images.
Since the total missing flux of the $^{12}$CO 12\,m+7\,m data in the observed field (Figure \ref{fig:COGMF} (a)) is $\sim$40\%, we additionally combined the IRAM image with the ALMA data using the \texttt{feathering} technique, and we confirmed that the combined image reproduces the total flux measured with the IRAM data alone. We assume that the $^{13}$CO, C$^{18}$O, and continuum observations have no significant missing flux, because the distributions of these tracers are more compact than those of $^{12}$CO. 

We retrieved the SUBARU H$\alpha$ Supreme-cam image (PI: Arimoto, N.; Proposal IDs: S01B091, S02B105) from the archive and calibrated with a standard manner to investigate the star formation activities in the GMC (see Sect.\,\ref{subs:COGMF}). 
The absolute astrometry of the original H$\alpha$ image is $\sim$0\farcs2, as we align the image using the UBNO-B1 catalog. To confirm the accuracy of the astrometry, a point-source catalog of Gaia Data Release 2 \citep{Bailer18} was used, and thus there is no significant positional error of the H$\alpha$ image over the ALMA beam size.  

\section{Results} \label{sec:results}
\subsection{Spatial Distributions of Molecular Gas of GMC-16} \label{subs:COGMF}
\begin{figure}[htbp]
\includegraphics[width=180mm]{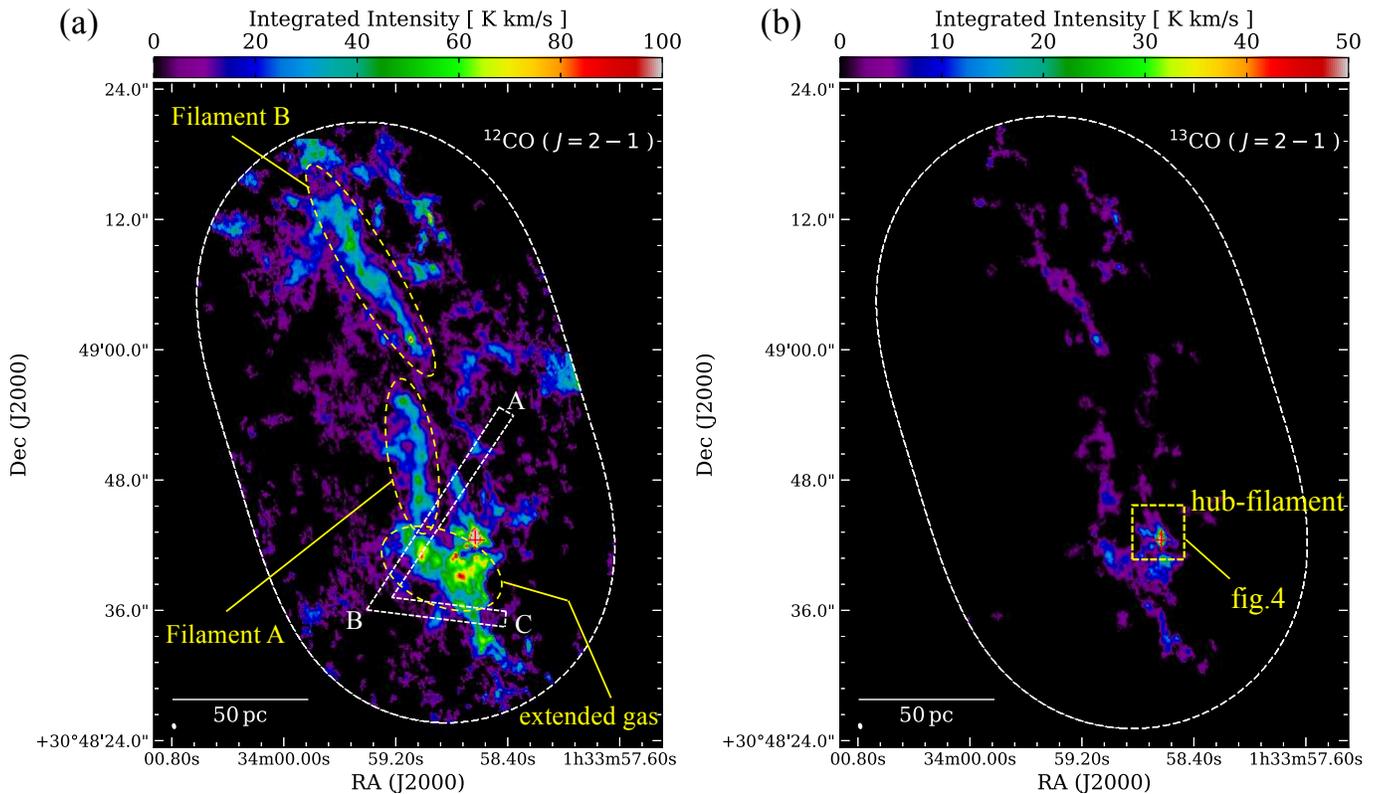}
\caption{Molecular gas distributions in M33-GMC-16. (a) Color-scale image shows the velocity-integrated intensity map of $^{12}$CO\,($J$\,=\,2--1) combining the ALMA data (the 12\,m + 7\,m array) with IRAM 30\,m data. The angular resolution is given by the white ellipse in the lower left corner. The cross mark denotes the position of the 1.3\,mm continuum peak.
Note that the color scale is adjusted to the range from 0 to 100\,K\,km\,s$^{-1}$ in order to show the diffuse emission and thus the CO peak is saturated in this figure. The white dotted line shows the field coverage of the ALMA observations. (b) Same as (a) but for the $^{13}$CO\,($J$\,=\,2--1) image obtained by the ALMA 12\,m + 7\,m array. 
\label{fig:COGMF}}
\end{figure}

Figure \ref{fig:COGMF} shows the molecular gas distributions in $^{12}$CO and $^{13}$CO\,($J$\,=\,2--1) in GMC-16. The missing flux of the 12\,m+7\,m data in $^{12}$CO toward the southern region where the decl. angle is lower than $+$30\arcdeg48\arcmin50\arcsec\ is as small as 10\%, indicating that the gas distributions are dominated by compact rather than the diffuse gas. This is consistent with an early result that active star-forming GMCs in the LMC are highly structured \citep{Sawada18}. The previous single-dish studies in CO lines by \cite{Tosaki11}, \cite{Miura12}, and \cite{Druard14} marginally resolved this GMC into two peaks. The ALMA observations clearly reveal multiple filamentary structures with the length scale of $\gtrsim$50\,pc elongated in the north--south direction. The filaments are not randomly distributed but exhibit ordered direction and there are roughly two main filamentary components (Filament A, and B) as indicated in Figure \ref{fig:COGMF} (a). In addition to these filaments, there is a relatively $``$extended gas$"$ (see the final paragraph of this subsection) and a dense clump associated with the 1.3\,mm continuum emission (hereafter, MMS, millimeter source) and multiple small filaments (see Sect.\,\ref{subs:outflow}) at the southern part of the GMC. 
We estimated the total gas mass and relatively dense gas mass from the $^{12}$CO and $^{13}$CO data, respectively. The first one is derived from the $^{12}$CO data assuming the $X_{\rm CO}$ factor in M33, 2.0\,$\times$\,10$^{20}$\,cm$^{-2}$\,(K\,km\,s$^{-1}$)$^{-1}$ \citep{Roso07}, and CO(2--1)/CO(1--0) ratio of 0.7 \citep{Tosaki11,Druard14}. The mass of higher density regions traced by $^{13}$CO is estimated by the local thermodynamical equilibrium (LTE) calculation applying the excitation temperature derived from the $^{12}$CO data and the relative abundance of [H$_{2}$]/[$^{13}$CO] of 1.4 $\times$ 10$^{6}$, which is close to an intermediate value adapted in the Galaxy \citep[e.g.,][]{Frerking82} and the LMC studies \citep[e.g.,][]{Fujii14,Fukui19,Tokuda19}.
The total mass ($^{12}$CO mass) and dense gas mass ($^{13}$CO mass) are 8 $\times$ 10$^{5}$\,$M_{\odot}$ and 2 $\times$ 10$^{5}$\,$M_{\odot}$, respectively. The fraction of $^{13}$CO/$^{12}$CO mass is $\sim$0.2, which is consistent with that in the Galactic plane \citep{Torii19}.
The total mass, median (FWHM) width, and length of Filament A/B in the $^{12}$CO map are measured to be $\sim$1 $\times$ 10$^{5}$\,$M_{\odot}$, $\sim$5--6\,pc, and $\sim$50--70\,pc, respectively. These characteristics are comparable to those of GMFs, e.g., the Nessie cloud \citep{Jackson10}, in the Galaxy. 

\begin{figure}[htbp]
\vspace{0.7cm}
\includegraphics[width=180mm]{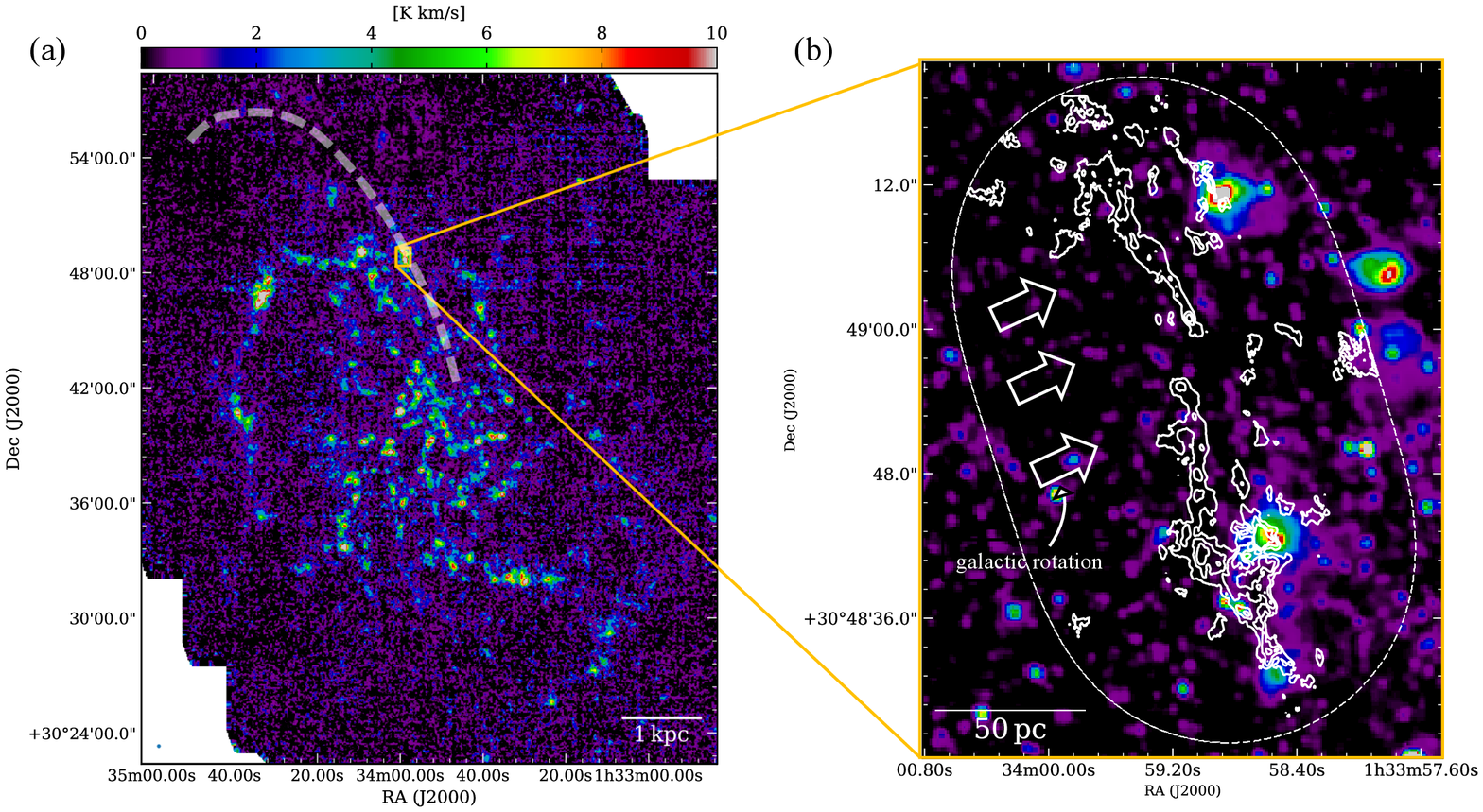}
\caption{(a) The $^{12}$CO\,($J$\,=\,2--1) velocity-integrated intensity map of M33 obtained with the IRAM 30\,m telescope \citep{Druard14}. The transparent white dashed line indicates the optical arm near GMC-16 \citep{Sandage80}. (b) Color-scale image shows H$\alpha$ emission obtained by the SUBARU telescope. Contour shows the $^{12}$CO map, which is the same as that in Figure \ref{fig:COGMF} (a). The lowest contour level and the subsequent steps are 20\,K\,km\,s$^{-1}$.
\label{fig:COHa}}
\end{figure}

Figure \ref{fig:COHa} (a) shows that GMC-16 is located at an optical spiral arm in the northern part of M33. Based on the morphology of the spiral arm, the moving direction of the gas (i.e., galactic rotation) is considered to be from east to west. Figure \ref{fig:COHa} (b) shows the comparison between the $^{12}$CO and H$\alpha$ images. There are several bright H$\;${\sc ii} regions and some of them are outside of the present field coverage with ALMA. Two GMFs (Filaments A, and B) are located at the eastern side of the H$\;${\sc ii} regions. Although the previous single-dish studies with a spatial resolution of a few $\times$10\,pc identified these H$\;${\sc ii} regions inside the molecular cloud, the present observations reveal a clear position offset between ionized and CO gas with a separation of $\sim$10--20\,pc.

\begin{figure}[htbp]
\begin{center}
\includegraphics[width=180mm]{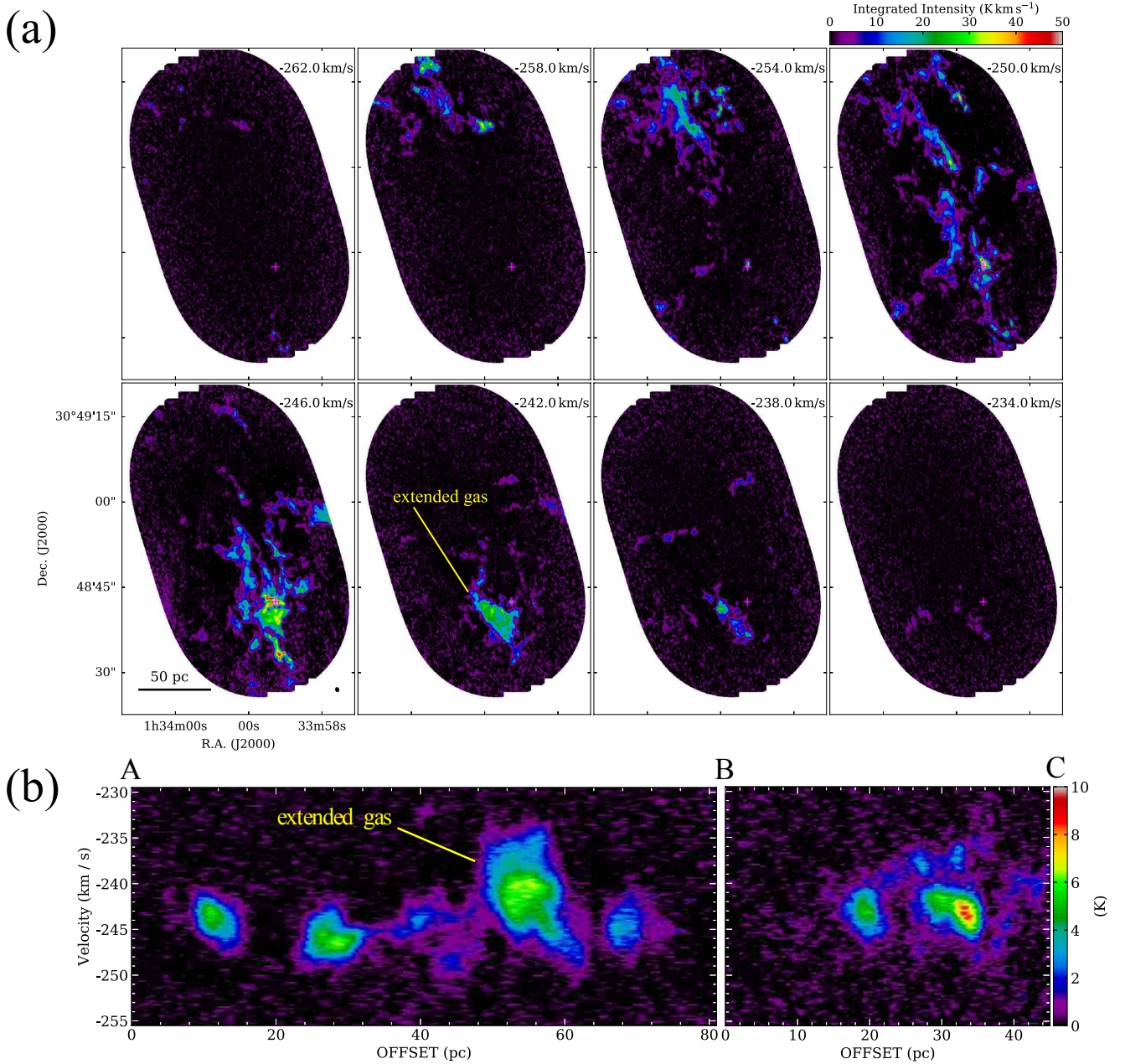}
\end{center}
\caption{(a) Velocity-channel maps toward GMC-16 in $^{12}$CO\,($J$\,=\,2--1). The lowest velocity is shown in the upper right corner in each panel. The angular resolution is given by the black ellipse in the lower right corner of the lower left panel. (b) A $^{12}$CO\,($J$\,=\,2--1) position-velocity diagram along the regions shown by the white dotted lines through three positions, A, B, and C in the Figure \ref{fig:COGMF} (a). 
\label{fig:COvel}}
\end{figure}

We made channel maps with a velocity bin of 4\,km\,s$^{-1}$ of the $^{12}$CO\,($J$\,=\,2--1) data to investigate the velocity structure of the GMC (Figure \ref{fig:COvel} (a)). As one can see in the channel maps, there is a velocity gradient along the north--south direction from blueshifted to redshifted velocity. 
To further illustrate the difference between the individual components, i.e., Filament A, extended gas, and some of the filaments, we made a position-velocity (PV) diagram of $^{12}$CO (Figure \ref{fig:COvel} (b)) along the dotted lines shown in Figure \ref{fig:COGMF} (a). We manually selected the analyzed region to avoid the complex hub-filamentary structure (Sect.\,\ref{subs:outflow}). The velocity width (FWHM) of the filamentary structures is measured as $\sim$3\,km\,s$^{-1}$, in contrast to this, the extended gas shows much wider velocity width, $\sim$6\,km\,s$^{-1}$. The centroid velocity of the extended gas is $\sim\!-$240\,km\,s$^{-1}$, which is apparently redshifted compared to that of Filament A. This difference may be relevant to turbulent dissipation and deceleration by the galactic spiral shock. We discuss this possibility in Sect.\,\ref{subs:disGMF}.

\subsection{A compact millimeter source with high-velocity wing components in GMC-16} \label{subs:outflow}
\begin{figure}[htbp]
\begin{center}
\includegraphics[width=170mm]{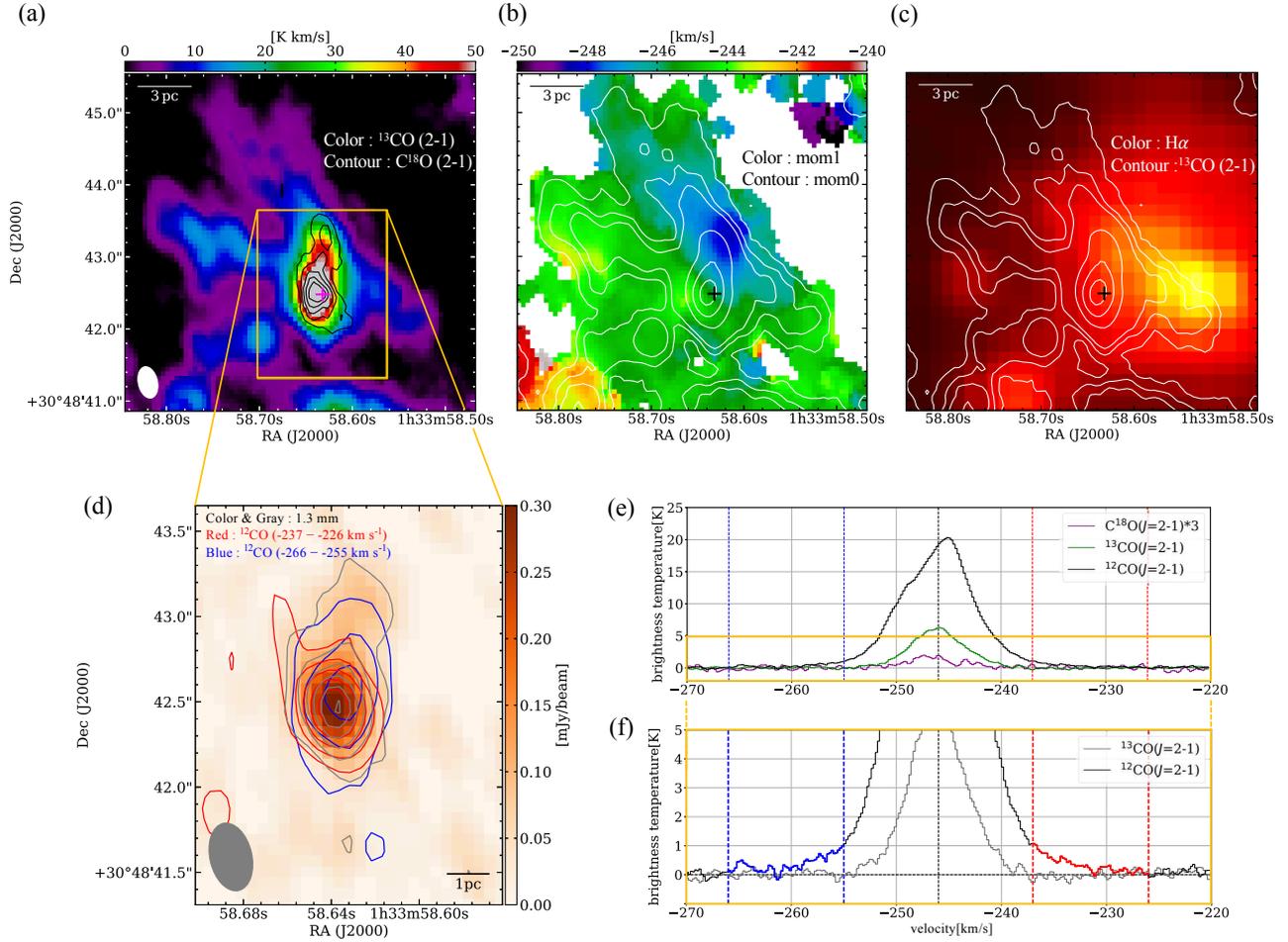}
\end{center}
\caption{Dense clump associated with the $^{12}$CO bipolar outflow candidate in GMC-16. (a) Color-scale and black contours show the $^{13}$CO and C$^{18}$O velocity-integrated intensity maps, respectively. The lowest contour level and the subsequent contour steps are 1.4\,K\,km\,s$^{-1}$. The cross denotes in panels (a), (b), and (c) the peak position of MMS shown in panel (d). The angular resolution is given by the white ellipse in the lower left corner. 
(b) Color-scale image and white contours show the $^{13}$CO moment\,1 map and moment\,0 map, respectively. The lowest contour and the subsequent steps are the 3$\sigma$ noise level, 2\,K\,km\,s$^{-1}$. 
(c) Color-scale image shows the H$\alpha$ distributions. White contours are the same as those in panel (b).
(d) Color-scale and gray contours show the 1.3\,mm continuum emission. The lowest contour level and the subsequent steps are the 3$\sigma$ noise level, 0.053\,mJy\,beam$^{-1}$. Red and blue contours represent the high-velocity $^{12}$CO emission originated from the outflowing gas. The integrated velocity ranges are shown by red and blue dotted lines in panels (e) and (f). The lowest contour and the subsequent steps are the 3$\sigma$ noise level, 3.3\,K\,km\,s$^{-1}$. 
(e) The $^{12}$CO, $^{13}$CO, and C$^{18}$O spectra at the peak position of MMS. Note that the $^{12}$CO data does not include the IRAM\,30m data. The black dotted line indicates the central velocity, $-$246\,km\,s$^{-1}$, obtained by fitting the $^{13}$CO spectra with a single Gaussian profile. The velocity ranges of red and blue dotted lines are $-$237 to $-$226\,km\,s$^{-1}$, and $-$266 to $-$255\,km\,s$^{-1}$, respectively.
(f) An enlarged view of the $^{12}$CO and $^{13}$CO profile in panel (e) to stress the wing feature of the outflowing gas. 
\label{fig:outflow}}
\end{figure}

Figure \ref{fig:outflow} shows a zoomed in view of the $^{12}$CO brightest peak clump in GMC-16. Panel (a) shows the $^{13}$CO distribution and there are several filamentary structures connecting to the central peak. This type of multiple filamentary structure, which is referred to as the $``$hub filament$"$ \citep{Myers09}, is found in high-mass star-forming regions in the Galaxy and the LMC \citep[e.g.,][]{Peretto13,Williams18,Fukui19,Tokuda19}. The hub filament in GMC-16 shows an asymmetric distribution, which is extended in a fan shape toward the east direction. The moment\,1 map in Figure \ref{fig:outflow} (b) shows that the northern filament has a central velocity of $\sim\!-$248\,km\,s$^{-1}$, which is different from those of the other filaments ($\sim\!-$245\,km\,s$^{-1}$). Note that the hub-filamentary complex is connected to the extended gas (see the moment\,0 map (Figure \ref{fig:COGMF}) in Sect.\,\ref{subs:COGMF}), but their central velocities are different from each other. 

We detected the 1.3\,mm continuum source (MMS) as well as the C$^{18}$O emission at the intersection of a few filaments, i.e., the $^{13}$CO peak (Figures \ref{fig:outflow} (a) and (d)). If there is a protostellar source inside MMS, the source may contribute to the enhancement of the 1.3\,mm flux. This effect is considered to be small, and the continuum flux is mainly arising from the cold dust emission, because the gas mass estimated from the 1.3\,mm continuum and that from the C$^{18}$O emission are similar to each other (see the next paragraph). For the same reason, it is likely that the free-free contamination from the H$\;${\sc ii} region next to MMS is also negligible. In fact, the peak position of the hub filament as well as MMS is shifted from that of the H$\alpha$ emission as shown in Figure \ref{fig:outflow} (c). The {\it Spitzer} observations found 24\,$\micron$ emission around this source with the intensity of 82\,mJy, which corresponds to the total infrared luminosity of $\sim$2 $\times$ 10$^{6}$\,$L_{\odot}$ \citep{Verley07}. Although the brightness of the source suggests the presence of early-O-type star(s) \citep{Martins05,Zinnecker07}, the luminosity is mainly arising from the high-mass stars inside the neighboring H$\;${\sc ii} region. 

The deconvolved (FWHM) size of MMS is (0\farcs53$\pm$0\farcs09) $\times$ (0\farcs27$\pm$0\farcs06), corresponding to $\sim$2.2\,pc $\times$ 1.1\,pc. 
The total mass of MMS ($M_{\rm total}^{MMS}$) above the 3$\sigma$ detection
is estimated to be $\sim$2 $\times$ 10$^{4}$\,$M_{\odot}$ assuming the dust opacity of $\kappa_{\rm 1.3 mm}$ of 1\,cm$^{2}$\,g$^{-1}$ for protostellar envelopes \citep[e.g.,][]{Oss94}, a dust-to-gas ratio of $\sim$3\,$\times$\,10$^{-3}$ (for the LMC-like metallicity; \citealt{Gordon14}), and the dust temperature of 20\,K. The velocity width (FWHM) of the C$^{18}$O emission in MMS is 6.8\,km\,s$^{-1}$. 
With a radius of 1.6\,pc (=geometric mean of the major and minor axis) the resultant virial mass, $\sim$2 $\times$ 10$^{4}$\,$M_{\odot}$, is consistent with the $M_{\rm total}^{MMS}$, indicating that the dense clump is gravitationally bound. Assuming a spherical geometry, the average density of MMS is calculated to be $\sim$2 $\times$ 10$^{4}$\,cm$^{-3}$. We note that we could not find any other continuum sources as well as C$^{18}$O emission in the observed field, indicating that the high-density region is localized within a few parsecs with respect to the entire molecular cloud of GMC-16.

Figure \ref{fig:outflow} (d) shows spatial distributions of the high-velocity $^{12}$CO gas toward MMS. The spectra of $^{12}$CO, $^{13}$CO, and C$^{18}$O are shown in panels (e) and (f). 
The $^{12}$CO profile is slightly asymmetric with a peak velocity of $\sim\!-$245\,km\,s$^{-1}$.  
As shown in the $^{13}$CO moment map in Figure \ref{fig:outflow} (b), there are two velocity components toward MMS, and the $^{12}$CO intensity of redshifted gas is stronger than that of the blueshifted gas. The double (or multiple) peaked profile in C$^{18}$O is insignificant at the present sensitivity. Since the $^{13}$CO profile shows a relatively symmetric shape, we use it to estimate the systemic velocity of MMS using Gaussian fitting. The systemic velocity judged from the $^{13}$CO profile is $\sim\!-$246\,km\,s$^{-1}$ and the maximum relative velocity of the $^{12}$CO red/blueshifted wing components is $\sim$20\,km\,s$^{-1}$. CO observations in the Galaxy and the LMC often found this type of high-velocity gas toward young stellar objects \citep[e.g.,][]{Beuther02,Fukui15,Fukui19,Shimonishi16,Harada19,Tokuda19}. 
The presence of the gravitationally bound dense material 
toward MMS strongly indicates that there is at least one embedded high-mass protostar and the $^{12}$CO high-velocity components are originated from its bipolar outflow. This is the first strong candidate for protostellar outflow in M33 as well as the external disk galaxies. We further discuss the reliability of the protostellar outflow in Sect.\,\ref{subs:Doutflow}.

\section{Discussions} \label{sec:dis}
\subsection{Possible origins of the GMFs} \label{subs:disGMF}
We discuss the formation mechanism of the GMFs in GMC-16. Early molecular gas surveys in the Galaxy speculated that this type of long filamentary structures is supposed to be formed by galactic spiral shocks \citep[e.g.,][]{Burton70,Jackson10}. Alternative ideas have also been proposed; for example, the streamer in W51 can be regarded as a part of the expansion ring based on its velocity structure \citep{Moon98}. Our extragalactic view of GMC-16 with a resolution similar to that of the Galactic single-dish studies allows us to address the origin of the GMFs. The morphology of the GMFs in GMC-16 shows a relatively straight shape, and the sizes of the H$\;${\sc ii} regions are small compared to those of the GMFs. Based on their morphologies, it is unlikely that the GMFs are regarded as a part of the ring-like structure associated with the H$\;${\sc ii} regions. 

As shown in Sect.\,\ref{subs:COGMF}, we find clear position discrepancies between the GMFs and the H$\;${\sc ii} regions. This type of spatial offset is sometimes seen in nearby grand design spiral galaxies \citep{Hirota11,Colombo14}, although the separations are more than a few $\times$10\,pc, which is much larger than the present case. The authors suggest that high-mass star formation activities are triggered by the density-wave driven galactic spiral shocks. The sequence of the GMFs and H$\;${\sc ii} regions in GMC-16 is consistent with the propagation direction of the spiral arm. According to the steady-state density-wave theory \citep{Lin64}, the interstellar medium is decelerated and compressed at the bottom of the stellar potential, i.e., the shock front \citep{Fujimoto68,Roberts69}. 
If we adopt a rotation velocity of the interstellar medium, $\sim$80\,km\,s$^{-1}$ at 2\,kpc from the galactic center \citep{Corbelli14}, and a pattern speed, $\Omega_{\rm p}$, of $\sim$25\,km\,s$^{-1}$\,\,kpc$^{-1}$ \citep{Newton80}, the timescale of the propagation of the H$\;${\sc ii} regions from the possible shock front traced by the CO filaments is calculated to be $\sim$1\,Myr (=20\,pc/ (80--50)\,km\,s$^{-1}$) assuming that the GMFs and the H$\;${\sc ii} regions are located at the same height in the galactic disk.
This is consistent with the age of the H$\;${\sc ii} regions judged from their sizes based on the Galactic studies \citep{Tremblin14}. 
This means that the two major H$\;${\sc ii} regions inside the observed field (Figure \ref{fig:COHa} (b)) might be formed around the present locations of Filament A/B, although the galactic shock alone may not be enough to trigger the high-mass star formation (see Sect.\,\ref{subs:dSF}).

We found a central velocity difference between the extended gas component ($\sim\!-$240\,km\,s$^{-1}$) at the eastern (upstream) side of GMC-16 and the filamentary structures ($\sim\!-$245\,km\,s$^{-1}$), as shown in Figure \ref{fig:COvel} and Sect.\,\ref{subs:COGMF}. Based on the $\sim$40\,pc resolution measurements in CO and H$\;${\sc i} by \cite{Gratier10}, the velocity gradient along the north--south direction of the GMC is $\lesssim$0.1\,km\,s$^{-1}$\,pc$^{-1}$, which is mostly arising from the galactic rotation. The rotational motion alone may not simply explain the velocity offset ($\sim$5\,km\,s$^{-1}$) in the PV diagram (Figure\,\ref{fig:COvel}(b)), which is extracted perpendicular to the large-scale gradient direction. In addition to this, the velocity width of the extended gas is larger than that of the filamentary clouds. These velocity features can be qualitatively explained by the dissipation of turbulent energy and the deceleration of the gas at the shock front. The idea that filamentary structure can be seen as a consequence of turbulent dissipation has been proposed in various scales of interstellar medium \citep[e.g.,][]{Padoan01,Inoue18,Tokuda18}. In addition to this, our observations toward GMC-8, which is a quiescent GMC, show fewer filamentary structures and a large velocity dispersion compared to the GMFs in GMC-16 (forthcoming paper). In summary, our indirect evidence suggests that the galactic spiral shock compresses a turbulent diffuse cloud to form filamentary structures. 

In flocculent spiral galaxies like M33, it may be hard to realize strong spiral shocks compared to grand design galaxies \citep{Egusa11,Egusa17,Hirota11,Colombo14}. \cite{Wada11} proposed that the complex spiral structures found in M33 are explained by the nonsteady stellar arms rather than the conventional density-wave picture. The galactic shear motion can elongate GMCs over an $\sim$100\,pc scale \citep{Wada08,Miyamoto14}. 
The nonsteady stellar arm model predicts that there is no clear spatial offset between the gas spiral arm and young stars across the galaxy. Further statistical studies to investigate the position discrepancies between molecular clouds and H$\;${\sc ii} regions based on observations toward similar targets in M33 with a similar angular resolution will allow us to draw a comprehensive picture of the relation between the galactic kinematics and GMF formation.

\subsection{ High-velocity wing components as a strong candidates for protostellar outflow} \label{subs:Doutflow}
We found the redshifted and blueshifted wing features at MMS (Sect.\,\ref{subs:outflow}). According to the Galactic studies of high-mass star-forming regions \citep[e.g.,][]{Beuther02}, such high-velocity wings with a maximum velocity of $\sim\!\pm$20\,km\,s$^{-1}$ or even much smaller velocities are normally interpreted as protostellar outflows. The typical size of outflow lobes in the Galaxy \citep[e.g.,][]{Beuther02} and the LMC \citep[e.g.,][]{Fukui19,Tokuda19} is less than 1\,pc. To demonstrate compactness of the possible outflow, we made the velocity profile maps  around the MMS'{}s location in $^{12}$CO and $^{13}$CO (Figure\,\ref{fig:pfmap} in the Appendix) with a grid size of 0\farcs93\,$\times$\,0\farcs61, which corresponds to $\sim$3.7\,pc\,$\times$\,$\sim$2.5\,pc. We could not find similar high-velocity wing emission except in the middle panel, i.e., at the location of MMS. These observational results strongly suggest that the high-velocity $^{12}$CO emission at MMS is a protostellar outflow.

We briefly mention the limitations of the present observations. As seen in Figure\,\ref{fig:outflow}, there is no significant spatial offset between the red and blue velocity components. Because protostellar outflows are as compact as $\lesssim$1\,pc as mentioned in the previous paragraph, this feature is mostly due to the lack of spatial resolution. For example, a low-resolution study toward the Galactic high-mass protostellar object AFGL\,2591 identified the molecular outflow without a clear offset between the red- and blueshifted components \citep{Lada84}. Subsequently, \cite{Mitchell91} clearly revealed its bipolar nature with a finer resolution measurement. 
A pole-on configuration is another possibility to interpret our data at MMS, in which case the accretion disk may be observable as face-on. However, the sizes of such disks associated with high-mass protostars are as small as $\lesssim$1,000\,au \cite[e.g.,][]{Motogi19}, which is much smaller than outflow lobes, and thus it is also impossible to resolve it with this measurement. The present elongation of MMS in the north--south direction is arising from the combination of the beam swelling effect and its filamentary nature.

Although we need much higher angular resolution observations to reveal the actual distribution of the outflow and its age, the current data tell us that the dynamical time (=size/velocity) is as young as $\sim$(5--8) $\times$ 10$^4$ yr depending on the assumption of the inclination angle (45\arcdeg--60\arcdeg). 
Considering the fact that the high-mass stars are frequently formed in binary or multiple manners \citep[e.g.,][]{Massey03}, MMS may be composed of unresolved multiple protostellar sources. Future high-resolution infrared and long-baseline ALMA observations will elucidate further details regarding the nature of the source. 

\subsection{A possible trigger of the high-mass star formation in GMC-16} \label{subs:dSF}
In Sect.\,\ref{subs:disGMF}, we suggest that the galactic shock may be a plausible mechanism to explain the GMF formation. However, the high-mass star-formation activity itself is localized within a few parsec scales rather than the entire $\sim$50\,pc scale filament. Although Filament B has a $^{13}$CO peak at the edge of the filament (Figure \ref{fig:COGMF}), we could not detect any dense gas tracers, C$^{18}$O, and 1.3\,mm continuum so far. With respect to the H$\;${\sc ii} regions, which represent the recent star formation activities (Figure \ref{fig:COHa} (b)), their distributions are discrete rather than continuous. These facts indicate that an additional factor may also be needed to trigger the high-mass star-formation. 

One possible mechanism to trigger the high-mass star formation in MMS is $``$collect and collapse$"$ \citep{Elmegreen77,Dale07} driven by the expansion of the nearby H$\;${\sc ii} region (Figure \ref{fig:outflow} (c)). However, the velocity gradient traced by the $^{13}$CO emission (Figure \ref{fig:outflow} (b)) does not follow the direction of the expanding motion judged from the distribution of the H$\;${\sc ii} region. It is unlikely that the second generation star formation in MMS of GMC-16 was triggered by the expanding motion. 

The moment\,1 map shows a drastic velocity change at the boundary between the northern and eastern filaments as mentioned in Sect.\,\ref{subs:outflow}. This type of velocity difference is often seen in high-mass star-forming filamentary clouds in the LMC at a subparsec resolution \citep{Fukui15,Saigo17,Harada19,Nayak19}, and its interpretation is that collision between two or several filaments trigger high-mass star formation. More recently, the follow-up observations toward some of the targets with a spatial resolution of $\lesssim$0.1\,pc resolved further complex substructures composed of many filaments, which are difficult to explain by a coalescence process of individual components \citep{Fukui19,Tokuda19}. 
Nevertheless, because the orientation of filaments in the two different high-mass star-forming regions separated by over 50\,pc is roughly aligned, they concluded that a tidally driven galactic scale colliding flow \citep[see][]{Fukui17,Tsuge19} induced the formation of the fan-shaped hub filaments as precursors of high-mass stars following the propagation direction of the flow.
Such a hub filament is also reproduced by numerical simulations of cloud--cloud collision \citep{Inoue18}.

In GMC-16, as shown in Figures \ref{fig:outflow} (a)--(c), the hub-filamentary structure extends toward the eastern direction, which is considered to be the upstream side in the spiral shock. Although the large-scale spiral shock may form the GMFs with a length of $\gtrsim$50\,pc as discussed in the Sect.\,\ref{subs:disGMF}, the formation of the small-scale hub filament may not be explained by the spiral shock alone. An additional factor, such as a collision between the GMF and a preexisting small cloud, is needed to interpret the localization of the hub-filamentary dense clump leading to high-mass star formation.
We note that a similar hub-like molecular complex is also found in NGC~604-GMC (K. Muraoka et al. 2020, in preparation). According to the galactic scale numerical simulations, cloud--cloud collisions frequently occur around the galactic potential where interstellar media are concentrated \cite[e.g.,][]{Dobbs14}. \cite{Sano19} found a high-mass star-forming clump possibly formed by a collision between two clouds in the central region in M33, suggesting that cloud--cloud collision is not a rare event in M33 (see also, \citealt{Tachihara18}; K. Muraoka et al. 2020, in preparation).
The Galactic high-mass star-forming regions NGC~6334 and NGC~6357 \citep{Persi08} are also good counterparts to consider the high-mass star formation activities analogous to GMC-16. The regions contain several bright infrared sources with a total luminosity of $>$10$^5$\,$L_{\odot}$ along the molecular filament with a length of $\sim$100\,pc. \cite{Fukui18} concluded that a cloud--cloud collision promoted over a 100\,pc scale mini-starburst in the NGC~6334 and NGC~6357 regions. Our GMC-16 studies may be equivalent to observations of such Galactic high-mass star-forming regions from outside the Galaxy.

\section{Summary} \label{sec:sum}
We have performed ALMA observations of one of the most massive GMCs in M33, GMC-16, associated with a few H$\;${\sc ii} regions. The spatial resolution is $\sim$1\,pc, which is the highest angular resolution molecular gas survey in M33 (see also K. Muraoka et al. 2020, in preparation). We have spatially resolved $>$50\,pc scale GMFs with a mass of $\sim$10$^{5}$\,$M_{\odot}$ along with the spiral arm in M33. One of the most striking features is a $\sim$10--20\,pc scale offset between the GMFs and H$\;${\sc ii} regions, suggesting that the density-wave driven galactic shock may convert the diffuse interstellar gas into the filamentary structures and promote the subsequent high-mass star formation. At the southern part of the GMC, the 1.3\,mm continuum and C$^{18}$O observations have found a $\sim$1\,pc scale dense clump (MMS) with an average number density of $\sim$10$^{4}$\,cm$^{-3}$ at the intersection of the hub-filamentary cloud possibly formed by a cloud--cloud collision. We found a promising candidate for protostellar outflow at one of the GMFs, indicating that high-mass star-formation is still ongoing in MMS.

\acknowledgments
This paper makes use of the following ALMA data: ADS/ JAO.ALMA\#2017.1.00461.S. ALMA is a partnership of the ESO, NSF, NINS, NRC, NSC, and ASIAA. The Joint ALMA Observatory is operated by the ESO, AUI/NRAO, and NAOJ. This work was supported by NAOJ ALMA Scientific Research grant Nos. 2016-03B and JSPS KAKENHI (grant Nos. 17K14251, 18K13580, 18K13582, 18K13587, 18H05440, 19K14758 and 19H05075). 
\software{CASA (v5.4.0; \citealt{McMullin07}), Astropy \citep{Astropy18}, APLpy \citep{Robi12}}

\begin{figure}[htbp]
\begin{center}
\includegraphics[width=180mm]{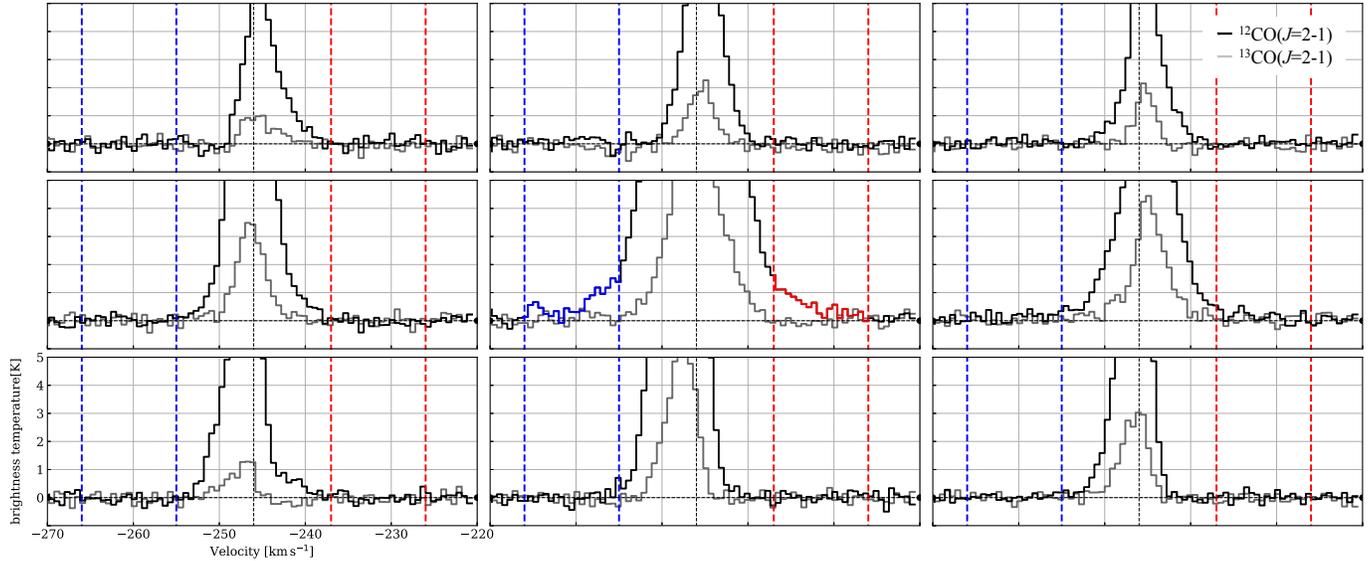}
\end{center}
\caption{Velocity profile maps in $^{12}$CO and $^{13}$CO centered at the position of MMS. The black, blue, and red dotted lines are the same as those in Figures\,\ref{fig:outflow} (e), (f).
\label{fig:pfmap}
}
\end{figure}

\appendix
Figure\,\ref{fig:pfmap} shows 3\,$\times$\,3 profile maps in $^{12}$CO and $^{13}$CO toward MMS. The map grid size of each panel is 0\farcs93\,$\times$\,0\farcs61, which corresponds to $\sim$3.7\,pc\,$\times$\,$\sim$2.5\,pc.


\bibliography{sample63}{}
\bibliographystyle{aasjournal}

\end{document}